\pdfoutput=1
\documentclass[journal,twoside,web]{ieeecolor2}
\usepackage{generic}
\usepackage{cite}
\usepackage{amsmath,amssymb,amsfonts}
\usepackage{algorithmic}
\usepackage{graphicx}
\usepackage{textcomp}
\usepackage{hyperref}
\hypersetup{hidelinks}
\usepackage{textcomp}
\usepackage{multirow}
\usepackage{booktabs}
\usepackage{colortbl}
\definecolor{mygray}{gray}{.85}
\def\BibTeX{{\rm B\kern-.05em{\sc i\kern-.025em b}\kern-.08em
    T\kern-.1667em\lower.7ex\hbox{E}\kern-.125emX}}
\begin{document}
\title{VIGIL: Vision-Language Guided Multiple Instance Learning Framework for Ulcerative Colitis Histological Healing Prediction}
\author{Zhengxuan Qiu, Bo Peng, Xiaoying Tang, \IEEEmembership{Senior Member, IEEE}, Jiankun Wang, \IEEEmembership{Senior Member, IEEE}, Qin Guo \vspace{-6mm}
\thanks{This work is partially supported by National Natural Science Foundation of China under Grant 62473191, Shenzhen Science and Technology Program under Grant  20231115141459001, Young Elite Scientists Sponsorship Program by CAST under Grant 2023QNRC001, High level of special funds (G03034K003) from Southern University of Science and Technology, Shenzhen, China. (\textit{Zhengxuan Qiu and Bo Peng contributed equally to this work) (Corresponding author: Jiankun Wang, Qin Guo)}}
\thanks{Zhengxuan Qiu, Xiaoying Tang and Jiankun Wang are with Shenzhen Key Laboratory of Robotics Perception and Intelligence and the Department of Electronic and Electrical Engineering at Southern University of Science and Technology in Shenzhen, China. (e-mail:  qiuzx2022@mail.sustech.edu.cn; tangxy@sustech.edu.cn; wangjk@sustech.edu.cn)}
\thanks{Jiankun Wang and Xiaoying Tang are also with the Jiaxing Research Institute, Southern University of Science and Technology in Jiaxing, China.}
\thanks{Bo Peng and Qin Guo are with the Department of Small Bowel Endoscopy, Department of Gastroenterology, Guangdong Provincial Key Laboratory of Colorectal and Pelvic Floor Diseases, and Biomedical Innovation Center, the Sixth Affiliated Hospital, Sun Yat-sen University in Guangzhou, China. (e-mail: pengb56@mail.sysu.edu.cn; guoq83@mail.sysu.edu.cn)}
}

\maketitle
\begin{abstract}

\textit{Objective: }Ulcerative colitis (UC), characterized by chronic inflammation with alternating remission-relapse cycles, requires precise histological healing (HH) evaluation to improve clinical outcomes. To overcome the limitations of annotation-intensive deep learning methods and suboptimal multi-instance learning (MIL) in HH prediction, we propose VIGIL, the first vision-language guided MIL framework integrating white light endoscopy (WLE) and endocytoscopy (EC).
\textit{Methods: }VIGIL begins with a dual-branch MIL module KS-MIL based on top-K typical frames selection and similarity metric adaptive learning to learn relationships among frame features effectively. By integrating the diagnostic report text and specially designed multi-level alignment and supervision between image-text pairs, VIGIL establishes joint image-text guidance during training to capture richer disease-related semantic information. Furthermore, VIGIL employs a multi-modal masked relation fusion (MMRF) strategy to uncover the latent diagnostic correlations of two endoscopic image representations. 
\textit{Results: }Comprehensive experiments on a real-world clinical dataset demonstrate VIGIL's superior performance, achieving 92.69\% accuracy and 94.79\% AUC, outperforming existing state-of-the-art methods.
\textit{Conclusion: }The proposed VIGIL framework successfully establishes an effective vision-language guided MIL paradigm for UC HH prediction, reducing annotation burdens while improving prediction reliability.
\textit{Significance: }The research outcomes provide new insights for non-invasive UC diagnosis and hold theoretical significance and clinical value for advancing intelligent healthcare development.

\end{abstract}

\begin{IEEEkeywords}
Ulcerative colitis, multiple instance learning, vision-language model, multi-modal fusion
\vspace{-1mm}
\end{IEEEkeywords}

\section{Introduction}
\label{sec:introduction}
\IEEEPARstart{U}{lcerative} colitis (UC) is an inflammatory bowel disease characterized by alternating remission and relapse. Patients often suffer from persistent or recurrent diarrhea, abdominal pain, gastrointestinal bleeding, and various systemic symptoms that can severely impact their quality of life. The exact pathogenesis of UC is still unclear, but it is generally linked to several factors, including genetics, environmental influences, immune system imbalances, and microbial infections\cite{kobayashi2020ulcerative}. Achieving histological healing (HH) is becoming an essential long-term treatment goal for UC, as it holds the promise of improving patients' future health outcomes\cite{narang2018association}. Currently, the evaluation of HH relies on pathology scoring of biopsy samples. However, this method has several drawbacks, including lengthy diagnostic times, high costs, risks of bleeding, and potential complications\cite{tontini2015advances}.

Examination techniques, such as endoscopy, play a crucial role in diagnosing UC. White light endoscopy (WLE) is commonly used to observe surface lesions of UC, revealing features such as fragile mucosa, exudates, ulcers, and erosions. Endocytoscopy (EC) is an emerging ultra-high magnification endoscopic technology that provides detailed identification of mucosal, vascular structures, and single-cell types through intraoperative staining, allowing for better non-invasive assessment of UC\cite{kudo2017classification}. At present, the diagnosis of UC via endoscopy relies heavily on expert clinicians manually reviewing the images, which is time-consuming, labor-intensive, and subjective. 

Deep learning technology is widely applied to various gastrointestinal lesion prediction tasks\cite{chadebecq2023artificial}\cite{iacucci2023artificial}. However, in actual clinical settings, patients' endoscopic videos often contain a large number of video frames, and acquiring accurate frame-level annotation can usually be challenging and resource-intensive. Furthermore, each frame may reflect the healing degree of UC, while in practice, doctors are more focused on the overall diagnostic outcomes for the patient. Among previous efforts to alleviate this problem, one promising approach is multiple instance learning (MIL), a sub-type of weakly supervised learning\cite{zhou2002neural}. By partitioning a patient's colonic segment sample (bag) into numerous video frames (instances), MIL establishes a model that integrates frame-level information for segment-level predictions, using only segment-level labels as weak supervision. MIL typically follows the instance, bag, or embedding-space paradigm to link the information between the frames and the overall colonic segment\cite{carbonneau2018multiple}\cite{del2022constrained}\cite{schwab2022automatic}.

\begin{figure}[t]
	\centering
	\includegraphics[width=\linewidth]{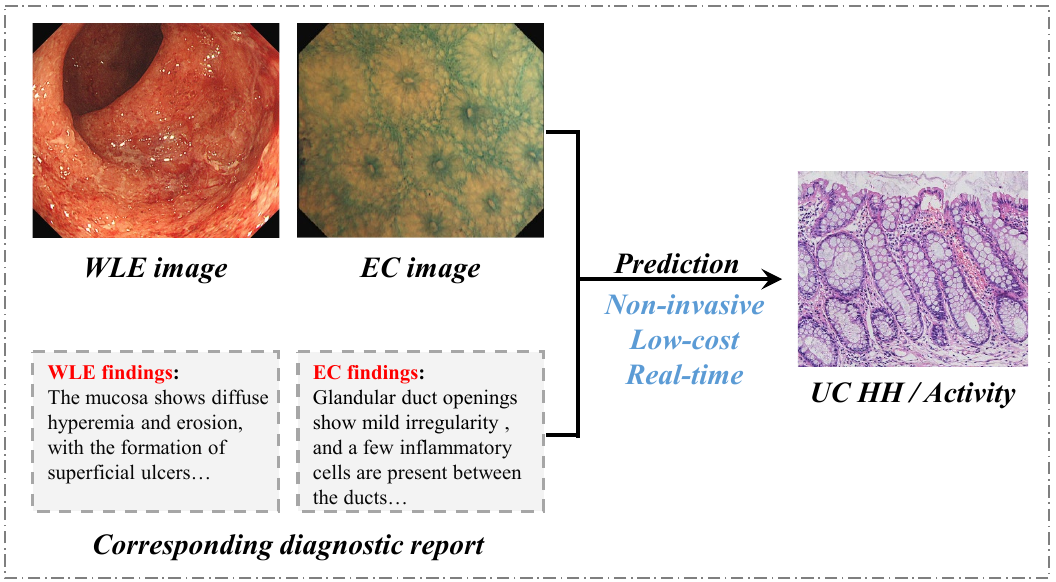}
	\caption{Illustration of UC HH prediction.}
	\label{ill}
        \vspace{-6mm}
\end{figure}

Practical clinical diagnosis and treatment rely not only on image information but also on other types of data, such as clinical data and diagnostic reports. With the rapid advancement of natural language processing (NLP) and multi-modal technologies, emerging research based on vision language models (VLM) such as CLIP\cite{radford2021clip} and BLIP\cite{li2022blip} has leveraged language information to provide new insights for the fields of image classification\cite{guo2023classification}, segmentation\cite{dhakal2024seg} and object detection\cite{kim2024detect}, etc. In the field of MIL, a series of VLM-based methods like MI-Zero\cite{lu2023MIzero}, PLIP\cite{huang2023vPLIP}, CONCH\cite{lu2024CONCH} and ViLa-MIL\cite{shi2024vila} have been introduced and achieved outstanding results across various medical image diagnostic tasks. However, in the area of endoscopic image, the potential to effectively harness VLM-based MIL for extracting valuable information as training constraints remains relatively unexplored, highlighting a promising avenue for further investigation.

The research above utilizing VLM primarily depends on image-text pair information. However, in the actual diagnostic process of UC, the integration of WLE and EC has demonstrated significant potential to improve the assessment of HH status\cite{neumann2013endocytoscopy}. The multi-modal fusion technique, which aims to integrate information representations from different modalities effectively, has been applied in various studies on the prediction of gastrointestinal disease\cite{ding2020scnet}\cite{lu2022real}\cite{chong2024cmmcan}. To the best of our knowledge,  there is currently a lack of research focused on the UC HH prediction based on multi-modal fusion. Integrating multi-scale information derived from WLE and EC is expected to offer a new perspective for UC HH prediction. 

Based on the considerations above, we propose a novel \textbf{V}\textbf{I}sion-language \textbf{G}uided multiple \textbf{I}nstance \textbf{L}earning framework (named VIGIL) for UC HH prediction task. Viewing the UC HH prediction as a colonic segment-frame-level MIL problem, we propose a dual-branch MIL feature integration module based on top-K typical frames and their similarity-weighted characteristics (named KS-MIL), which effectively explores the correlations among instance features. Subsequently, mirroring the imaging assessment habits of clinicians, we approach multi-modal information processing from two perspectives: for image-text pairs, we designed a sub-framework inspired by CLIP, employing multi-level contrastive learning to achieve effective alignment between images and texts; for the two types of images, WLE and EC, our proposed multi-modal masked relation fusion (MMRF) strategy integrates the surface lesions and cellular-level changes represented by both images, capturing potential latent feature relationships within the data, thereby providing new evidence for UC HH prediction. Specifically, the main contributions of this paper can be summarized as follows:

\begin{itemize}
    \item A vision-language guided MIL framework is proposed, establishing the information correlations among three aspects: colonic segment-frame level, image-text, and image-image (WLE and EC).  
    \item A dual-branch MIL module is designed with top-K typical instances selection and similarity metric learning. Discriminative feature integration is enhanced through inter-instance correlation analysis, improving UC healing status identification.
    \item A CLIP-based vision-language sub-framework is developed, incorporating a multi-level alignment and supervision strategy (global sample-level alignment, local token-level alignment, and disease-level supervision) to learn generalized and discriminative visual representation for the downstream classification task.
    \item An MMRF strategy is introduced for WLE-EC fusion, employing intra-/cross-modal self-attention mechanisms with relational masking to extract lesion-specific correlations across modalities.
\end{itemize}


\section{Related Works}
\subsection{Deep learning-based UC HH prediction frameworks}
In recent years, many researchers have proposed deep learning methods for UC HH prediction. Most of these studies utilized WLE or pathology images as input for the models. Ozawa \textit{et al}. constructed a computer-assisted diagnosis system based on GoogLeNet to identify endoscopic inflammation severity in patients with UC\cite{ozawa2019novel}. Sutton \textit{et al}. utilized a pretrained DenseNet121 architecture along with the grid search technique to identify the best hyperparameters for diagnosis of UC in endoscopy images\cite{sutton2022artificial}. Iacucci \textit{et al}. developed a CNN model based on VGG16 to predict healing or activity of UC in pathology images\cite{iacucci2023artificial}. An attention-constrained module was implemented to focus on the regions of neutrophils, facilitated by additional detailed pixel-level annotations, increasing the manual workload. 

While these studies demonstrate great classification performance in UC, they face several limitations. Firstly, from a clinical perspective, healing observed under WLE may not fully reflect HH status due to potential persistent histological inflammation\cite{nardone2019can}\cite{bryant2016beyond}. Additionally, acquiring pathology images through biopsy can introduce complications and other issues. The introduction of EC has made it possible to perform non-invasive pathology examinations in UC\cite{nakazato2017endocytoscopy}. Moreover, these studies require detailed annotations for each endoscopic or pathology image, making the real-world application impractical. Furthermore, they primarily analyze data based on a single image modality, while the factors influencing HH in UC are complex and diverse. If complementary information from multiple modalities could be further utilized, it would provide a more comprehensive understanding of HH prediction in UC.

\vspace{-4.5mm}
\subsection{Multiple Instance Learning}
MIL, as a common weakly supervised image classification method, can alleviate the burden of detailed annotations to some extent. Some researchers have applied the MIL approach to UC HH prediction. Schwab \textit{et al}. developed a novel ordinal MIL method for UC severity prediction\cite{schwab2022automatic}. The method utilized three CNNs to predict severity scores for each frame, which were then aggregated using maximum, threshold, and summation operators to derive the overall UC severity score. Del Amor \textit{et al}. proposed a MIL model for UC prediction using pathology images\cite{del2022constrained}. This model used a weighted average of instances where weights were obtained from supervised activation maps. ABMIL\cite{ilse2018attention} may be the most popular MIL approach, which employs a self-attention aggregation module to learn the weights of individual instances using a parameterized neural network. Based on ABMIL, a series of approaches has been proposed to explore how to integrate instance features more effectively. DSMIL designed a dual-stream architecture based on max-pooling and attention mechanism\cite{li2021dsmil}. TransMIL\cite{shao2021transmil} adopted a transformer architecture that explores morphological and spatial information between instances. PMIL\cite{yu2023prototypical} introduced prototype-based MIL into computational pathology, where prototypes were acquired from training data by unsupervised clustering. However, this method is highly dependent on the clustering quality and is noise-sensitive. In this work, we propose a dual-branch MIL module based on top-K typical instances and similarity metric adaptive learning to enhance the integration of discriminative instance features effectively.

The vision-language models, such as CLIP\cite{radford2021clip} and BLIP\cite{li2022blip}, have shown remarkable performance on a range of downstream visual recognition tasks. CLIP-like models utilize a dual-encoder image-text architecture, which are pre-trained on a large number of web-source datasets by exploiting the cross-modal alignment between paired images and captions. In the field of MIL, recent researchers have explored how to improve the sample efficiency and model performance based on CLIP. MI-Zero\cite{lu2023MIzero} employed pre-trained CTransPath\cite{wang2022ctrans} to extract pathology image patch features and utilized top-K pooling to integrate multiple instance features. The model has undergone pre-training on a dataset comprising 33,000 image-caption pairs. Later, PLIP\cite{huang2023vPLIP} built an extensive dataset consisting of 208,414 pathology images paired with natural language descriptions and fine-tuned a pre-trained CLIP model using this dataset. CONCH\cite{lu2024CONCH} fine-tuned the SOTA visual-language foundation pre-training framework CoCa\cite{yu2022coca} with a dataset of over 1.17 million image-caption pairs sourced from various origins. Although these studies have achieved impressive classification performance and transferability, building such a large number of image-caption pairs remains labor-intensive and time-consuming. Some VLM-based MIL methods have recently attempted to explore parameter-efficient methods for fine-tuning CLIP models. For example, ViLa-MIL\cite{shi2024vila} designed a dual-scale VLM-based MIL framework for pathology image classification by introducing two trainable prototype-based decoders for image and text branches to adapt the VLM for processing the image efficiently. However, it still faces challenges similar to prototype-based MIL methods and relies on prevalent global sample contrastive learning for cross-modal alignment, which may not fully leverage valuable local features. In this work, our VIGIL introduces a vision-language sub-framework and utilizes a multi-level contrastive learning strategy for comprehensive image-text alignment.

\vspace{-3mm}
\subsection{Muti-modal Fusion in Endoscopic Image}
Multi-modal fusion aims to integrate information from different modalities while preserving the unique semantic information of each modality, minimizing the heterogeneity between modalities, and ultimately achieving more effective information representation for downstream tasks\cite{baltruvsaitis2018multimodal}. Recently, some researchers have applied multi-modal fusion to the field of gastrointestinal disease based on endoscopic images, yielding promising results. Zhu \textit{et al.} developed a multi-modal fusion framework based on CNN and LSTM to explore the diagnosis of gastric protruded lesions using WLE and endoscopic ultrasound images as input. Endo-CRC \cite{lu2022real} employed a self-constructed CNN to integrate concatenated features derived from WLE and image-enhanced endoscopy images, facilitating efficient diagnosis of colorectal cancer. CMMCAN \cite{chong2024cmmcan} developed a multi-modal feature fusion module utilizing an adaptive attention mechanism to examine the correlation between endoscopic grayscale and optical flow images. All of these methods still rely on simple concatenation or cross-modal attention for multi-modal fusion, so further investigation is needed to fully capture the intricate relationships between modalities and effectively leverage the distinctive semantic information inherent within each modality. In this work, we introduce an MMRF strategy to leverage cross-modal disease associations while preserving modality-specific semantic information, thereby achieving more accurate and reliable UC HH prediction.

\section{Methodology}
\begin{figure*}[!t]
	\centering
	\includegraphics[width=0.9\linewidth]{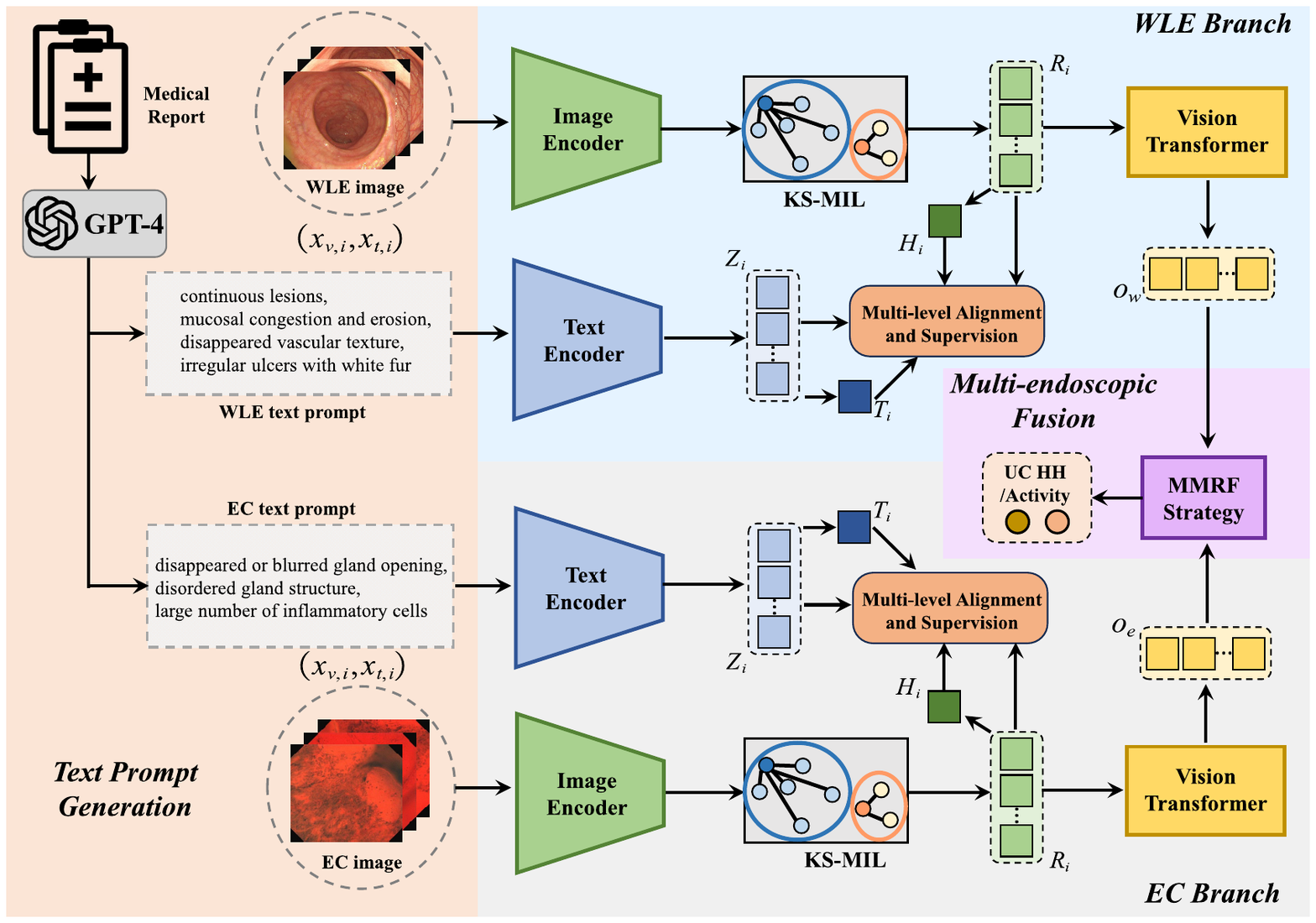}
	\caption{Overview of the proposed VIGIL framework. The VIGIL model primarily comprises following parts: text prompt generation, KS-MIL, vision-language sub-framework (WLE and EC branch), and multi-endoscopic fusion.}
	\label{overview}
        \vspace{-5mm}
\end{figure*}

In this paper, a novel vision-language guided MIL framework, VIGIL, is proposed (Fig. \ref{overview}). Firstly, by viewing UC HH prediction as a MIL problem, we describe our problem statement. We then present the top-K similarity-weighted (KS-MIL) module for effective segment-frame level prediction. Moreover, a vision-language sub-framework and a multi-level alignment and supervision training strategy are introduced, incorporating text prompts derived from diagnostic reports to boost diagnostic performance. Finally, the multi-modal masked relation fusion (MMRF) strategy for combining WLE and EC is presented in the multi-endoscopic fusion stage.

\vspace{-3mm}
\subsection{MIL Problem Statement}
We formulate UC colonic segment-level prediction as an MIL problem, where a segment (bag) consists of multiple video frames (instances). Crucially, MIL typically requires only bag-level labels, obviating the need for instance-level annotations, thus reducing annotation effort. For binary classification, given a bag $B=\left\{\left(x_{1}, y_{1}\right), \ldots,\left(x_{n}, y_{n}\right)\right\}$ of instances $x_{i} \in \mathcal{X}$ with corresponding labels $y_{i} \in\{0,1\}$, where $y_{i}=1$ indicates a segment in UC activity and $y_{i}=0$ indicates a segment in UC HH status, the bag label $c(B)$ is defined as:
\begin{equation}
    c(B)=\left\{\begin{array}{ll}
    0, & \text { if } \sum y_{i}=0 \\
    1, & \text { otherwise }\end{array}\right.
\end{equation}

Our task aims to predict the bag label $c(B)$ through operations on the instances that predict the UC HH status from input WLE and EC. The key challenge in MIL lies in establishing the connection between instance-level features and bag-level predictions. Existing MIL methods generally fall under two paradigms: instance-space (IS) and embedding-space (ES), expressed respectively as:
\begin{equation}
    c_{IS}(B) = g[f(\hat{x_1}), f(\hat{x_2}), \ldots, f(\hat{x_n})]
\end{equation}
\begin{equation}
    c_{ES}(B) = f[g(\hat{x_1}, \hat{x_2}, \ldots, \hat{x_n})]
\end{equation}
where $\hat{x_n}$ denotes the feature extracted from the instance $x_n$, $f$ and $g$ refer to the operations of classification and aggregation, respectively.

\vspace{-3mm}
\subsection{Proposed Top-K Similarity-weighted (KS-MIL) Module}

\begin{figure}[!t]
	\centering
	\includegraphics[width=\linewidth]{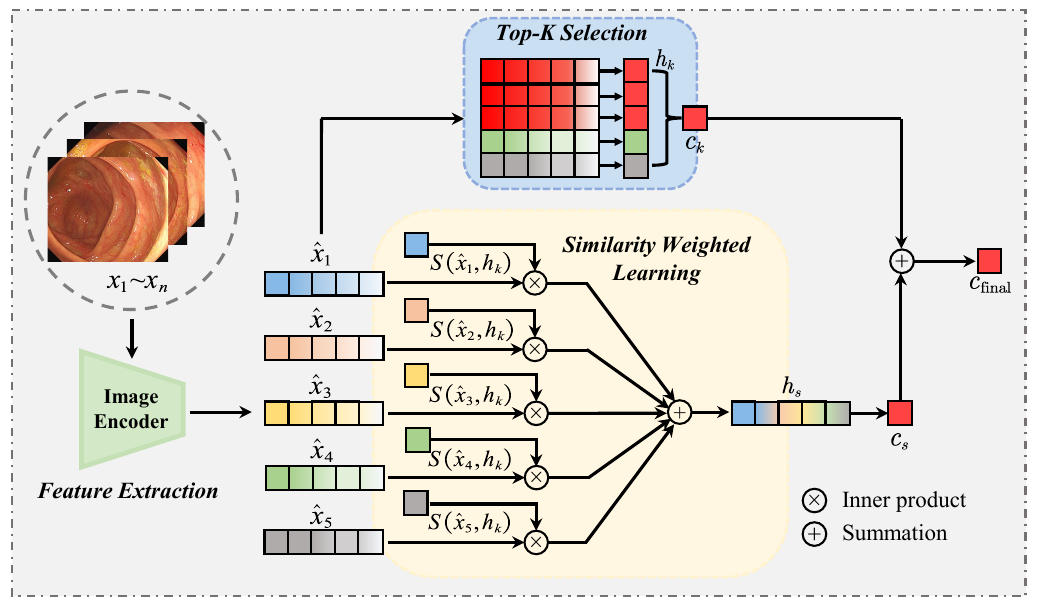}
	\caption{Detailed structure of the proposed KS-MIL, including two parts: top-K typical frames selection and similarity weighted metric learning. }
	\label{ksmil}
        \vspace{-5mm}
\end{figure}

Mirroring the diagnostic process of clinicians, typical frames are often selected from the video to reflect the disease status. Furthermore, the relationship between these individual frames and key typical frames must be considered for a more comprehensive prediction. Given these considerations, we propose KS-MIL in Fig. \ref{ksmil} that consists of a top-K selector branch and a similarity metric learning branch for feature aggregation. Unlike most existing methods that learn either instance or bag classifiers independently, KS-MIL jointly learns both, along with the bag embedding, using a dual-branch architecture.

The first branch is designed to select top-K typical frames, adheres to the instance-space paradigm. Let $\hat{x_i}$ denotes the feature embeddings extracted from instance $x_i$ using a feature extractor (typically a CNN). To be specific, the top-K selector employs a parameterized fully-connected layer with a sigmoid activation to score each instance feature $\hat{x_i}$ based on its relevance to the UC activity class: $p_i = \text{Sigmoid}(\hat{x_i})$. Subsequently, the instance features are sorted in descending order based on their calculated scores $\{p_1, p_2, \ldots, p_n\}$, indexed by 
$I_{K}=\left\{n_{1}, \ldots, n_{K}\right\}$, then the top-K features are mean-pooled to produce the aggregated typical feature embedding $h_K$ for this branch, as defined by:\
\begin{equation}
    h_k =\frac{1}{K} \sum_{i \in I_{K}} \hat{x_i}
\end{equation}
The prediction score of the bag $c_K$ is:
\begin{equation}
    c_k = W_kh_k
\end{equation}

The second branch is developed to measure the similarity between each instance with typical frame, adheres to the embedding-space paradigm. Inspired by the self-attention mechanism\cite{ilse2018attention}, the obtained typical feature embedding $c_K$ and each instance $\hat{x_i}$ undergo linear projections to produce two vectors, query $q_i$ and value $v_i$. For example, the projection formula for $\hat{x_i}$ is:
\begin{equation}
    q_i = W_q\hat{x_i}, v_i = W_v\hat{x_i}
\end{equation}
where $W_q$ and $W_v$ are weight matrices. Then we introduce a metric measurement $S$ based on cosine similarity to quantify the similarity score between an arbitrary instance and the typical instance as:
\begin{equation}
    S(\hat{x_i}, h_k) = \text{Softmax}(\frac{<q_i, q_K>}{\sqrt{d_q}})
\end{equation}
where $<.,.>$ refers to the inner product of two vectors. $d_q$ denotes the dimension of the query.
Therefore, the bag-level embedding $h_s$ is aggregated as a weighted sum of instance embeddings, where the weights are the above similarity scores $S$:
\begin{equation}
    h_s = \sum_{i=1}^{n} S(\hat{x_i}, h_k)v_i
\end{equation}
The prediction bag score of the second branch $c_s$ is:
\begin{equation}
    c_s = W_sh_s 
\end{equation}

The similarity between queries is measured using cosine similarity. Instances more similar to the typical instance are assigned greater weights to enhance bag-level prediction by providing more representative or disease-relevant features. Concurrently, an additional value vector $v_i$ ensures that the contributing information of the intra-instance is preserved while an inter-instance interaction is achieved by similarity to the typical instance.

Finally, the final prediction score for the bag is derived by integrating the scores from the two branches described above:
\begin{equation}
    c_{\text{final}} = \frac{1}{2}(c_k+c_s)
\end{equation}
and the final bag embedding $h_{\text{final}} = h_s$.

\vspace{-3mm}
\subsection{Vision-language Sub-framework}
Recent advances in MIL have taken advantage of VLM, particularly those based on CLIP\cite{radford2021clip}, to achieve notable results\cite{lu2023MIzero}\cite{huang2023vPLIP}\cite{shi2024vila}. Building on this success, we propose a vision-language sub-framework grounded in CLIP to further explore and elucidate the latent relationship with image-text pairs. Globally, it employs a dual-tower architecture comprising an image branch and a text branch. To obtain disease-relevant representations of samples, the architecture involves two key components: 1) Text prompt generation and tokenization; 2) Multi-level alignment and supervision.
\subsubsection{Text Prompt Generation and Tokenization}
To enhance the effectiveness of CLIP-based VLM, one of the most critical challenges is how to construct text prompts that accurately capture the pathological states depicted in WLE or EC. We collect diagnostic reports associated with WLE and EC from real-world clinical settings. These reports contain textual descriptions of pathological features, including inflammation, exudates, bleeding, and erosions observed in WLE, as well as cellular infiltration, glandular gaps, and openings visualized in EC. Leveraging this a prior knowledge, we employ text prompts derived from these diagnostic reports to guide VLM for UC HH prediction.

To address redundancy and irrelevant information in long texts, as well as input length constraints of text encoders, we first employ GPT-4 to automatically extract disease-relevant text prompts from the original diagnostic reports. Specifically, by prompting GPT-4 with specific instructions, lists of disease-relevant text prompts can be generated, \textit{e.g.}, \textit{"WLE: continuous lesions, fine granular mucosal changes, congestion and erosion, disappeared vascular texture", "EC: uneven gland density, branching gland openings, many inflammatory cells infiltration."} These text prompts have been validated by clinicians to ensure they are correct. The generation of text prompts via GPT-4 is performed offline as a pre-processing step.

A training set $\mathcal{D}$ consists of $N$ image-text pairs is defined as: $\mathcal{D}=\left\{\left(x_{v, 1}, x_{t, 1}\right),\left(x_{v, 2}, x_{t, 2}\right), \ldots,\left(x_{v, N}, x_{t, N}\right)\right\}$ where $x_{v,i}$ is the segment that contains several video frames, $x_{t,i}$ is the corresponding generated text prompt. For the $i$-th image-text pair $\left(x_{v, i}, x_{t, i}\right)$, a text encoder DistilBERT\cite{sanh2019distilbert} generates encoded text tokens $Z_i = \{z_{i,1}, z_{i,2}, \ldots , z_{i,L}\}$ and a global text representation $T_i$, while the image encoder based on proposed KS-MIL generates encoded visual tokens $R_i = \{r_{i,1}, r_{i,2}, \ldots , r_{i,M}\}$ and a global visual representation $H_i$. 

\subsubsection{Multi-level Alignment and Supervision}

As illustrated in Fig. \ref{multi}, we introduce a multi-level training strategy consisting of global sample-level alignment, local token-level alignment, and disease-level supervision for representation learning between endoscopic images and diagnostic reports.

The global sample-level alignment seeks to closely align representations of matching image-text pairs in the latent space while mapping mismatched pairs far apart. Following well-known practice\cite{radford2021clip}, the image and text global representations $H_i$ and $T_i$ are projected into normalized embeddings $\hat{H_i}$ and $\hat{T_i}$ via two independent non-linear projection layers. We leverage the InfoNCE\cite{oord2018infonce} loss to maximize the mutual information between paired image and text representations, thereby explicitly enforcing alignment at the global level. In detail, the global sample-level alignment loss $L_{\text{global}}$ is denoted as:
\begin{equation}
    L_g^{v\rightarrow{t}} = -\log \frac{\exp(\hat{H_i}^T\hat{T_i}/\tau_1)}
    {\sum_{k=1}^B\exp(\hat{H_i}^T\hat{T_k}/\tau_1)}
\end{equation}
\begin{equation}
    L_g^{t\rightarrow{v}} = -\log \frac{\exp(\hat{T_i}^T\hat{H_i}/\tau_1)}
    {\sum_{k=1}^B\exp(\hat{T_i}^T\hat{H_k}/\tau_1)}
\end{equation}
\begin{equation}
    L_{\text{global}} = \frac{1}{2N}\sum_{i=1}^N(L_g^{v\rightarrow{t}}+L_g^{t\rightarrow{v}})
\end{equation}
where $B$ is the batch size, $\tau_1$ is the temperature hyperparameter and $N$ is the number of image-text pairs.

In medical images, lesions often occupy only a small portion of the image, and correspondingly, only a few disease-related tags within the associated text report describe the key diagnostic findings. As such, we employ the local token-level alignment based on cross attention mechanism to explicitly match and align local representation between image-text pairs. Specifically, the encoded visual and text tokens $R_i$ and $Z_i$ are projected into normalized embeddings, which result in $\hat{R_i} = \{\hat{r}_{i,1}, \hat{r}_{i,2}, \ldots , \hat{r}_{i,M}\}$ and $\hat{Z_i} = \{\hat{z}_{i,1}, \hat{z}_{i,2}, \ldots , \hat{z}_{i,L}\}$. Then we need to find the soft matching between visual and text tokens with cross-attention mechanism\cite{chen2020uniter}\cite{lu2016hierarchical}. For the $j$-th visual token embedding $\hat{r}_{i,j}$ in $i$-th image-text pair,  $\hat{r}_{i,j}$ is attended to all text token embeddings in $\hat{Z_i}$, and the corresponding cross-modal text embedding $\widetilde{Z}_{i,j}$ can be formulated as:
\begin{equation}
    \widetilde{Z}_{i,j} = \sum_{k=1}^N\text{Softmax}(\frac{(Q\hat{r}_{i,j}) (K\hat{z}_{i,k})^T}{\sqrt{d}})(V\hat{z}_{i,k})
\end{equation}
where $Q,K,V$ are learnable matrices and $d$ is the dimension of these matrices.
Lying in the purpose of maximizing the lower bound of mutual information, we also utilize InfoNCE loss to pull $\hat{r}_{i,j}$ close to corresponding $\widetilde{Z}_{i,j}$ but push it away from other cross-modal text embeddings:

\begin{figure}[!t]
	\centering
	\includegraphics[width=0.88\linewidth]{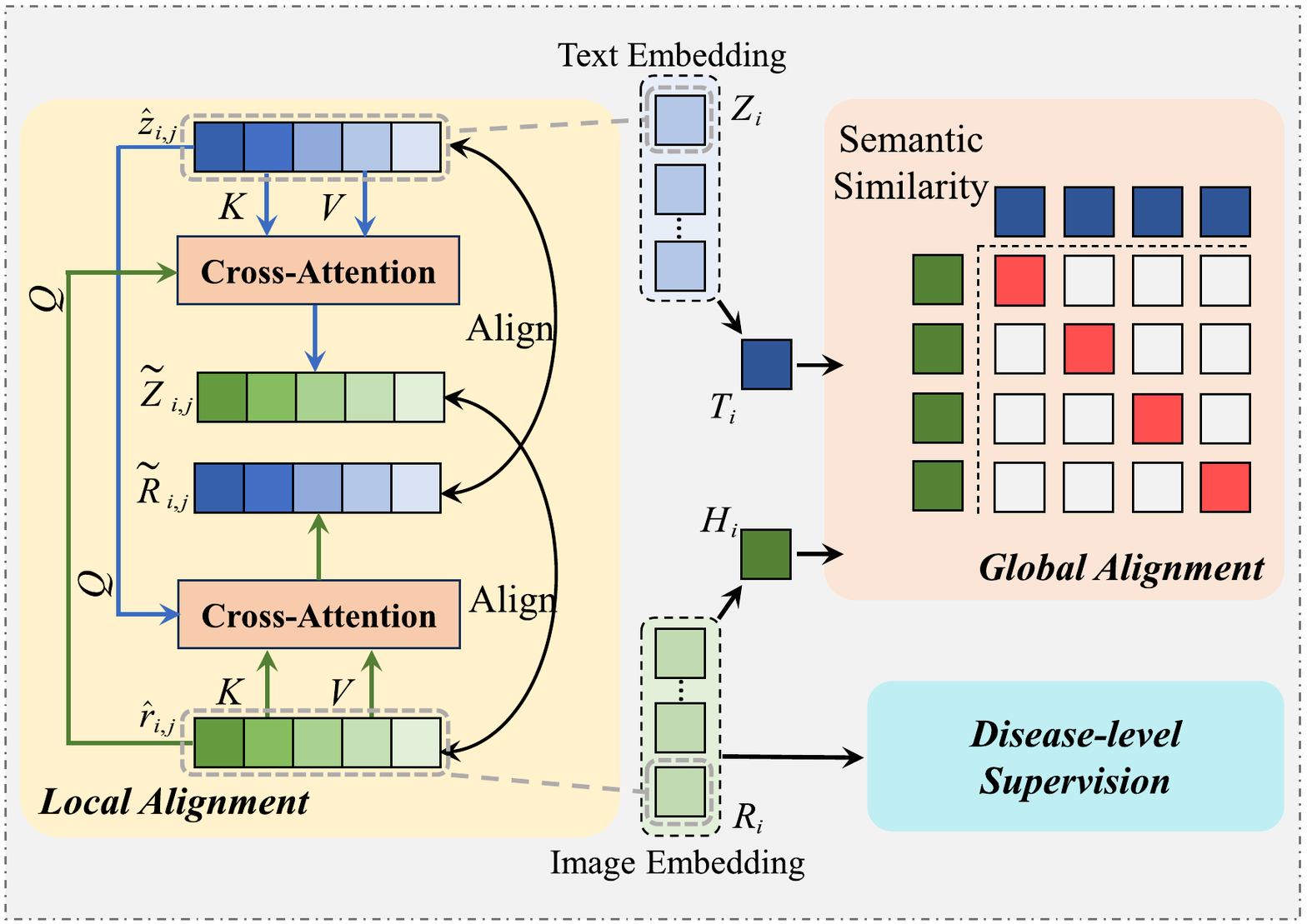}
	\caption{Detailed structure of multi-level alignment and supervision, which consists of global sample-level alignment, local token-level
alignment, and disease-level supervision. }
	\label{multi}
        \vspace{-6mm}
\end{figure}

\begin{align}
    L_l^{v\rightarrow{t}} = &-\frac{1}{2NM}\sum_{i=1}^N \sum_{j=1}^M w_{i,j} (
    \log \frac{\exp((\hat{r}_{i,j})^T\widetilde{Z}_{i,j}/\tau_2)}{\sum_{k=1}^M\exp((\hat{r}_{i,j})^T\widetilde{Z}_{i,k}/\tau_2)}\notag\\
    &+ 
    \log \frac{\exp({(\widetilde{Z}_{i,j})}^T\hat{r}_{i,j}/\tau_2)}{\sum_{k=1}^M\exp((\widetilde{Z}_{i,j})^T\hat{r}_{i,k}/\tau_2)}
    )
\end{align}
where $w_{i,j}$ is a learnable weight matrix and $\tau_2$ is the temperature hyperparameter.

Similarly, for the $j$-th text token embedding $\hat{z}_{i,j}$, we can calculate its corresponding cross-modal image embedding $\widetilde{R}_{i,j}$ and establish the loss function $L_l^{t\rightarrow{v}}$ by contrasting $\hat{z}_{i,j}$ with $\widetilde{R}_{i,j}$. Therefore, the local token-level alignment loss $L_\text{local}$ is given by:
\begin{equation}
    L_{\text{local}} = \frac{1}{2} (L_l^{v\rightarrow{t}}+L_l^{t\rightarrow{v}})
\end{equation}

Notably, the text input is employed only during training phase to provide additional supervision and guide the image branch towards extracting more disease-relevant semantic features. During inference phase, only the test image is required as input. Furthermore, to accurately predict UC HH status, the aligned global and local image embeddings are concatenated and then fed into a Vision Transformer (VIT) encoder\cite{dosovitskiy2020vit} to obtain disease-relevant representations. Finally, a parameterized fully-connected layer $F_y$ generates a prediction for the sample. The model is trained by minimizing the cross-entropy loss between the predicted and true probability distributions, providing disease-level supervision: 
\begin{equation}
    L_{\text{disease}} = -\frac{1}{N}\sum_{i=1}^N y_i \log(F_y(\text{VIT}(\hat{H_i}\oplus\hat{R_i})))
\end{equation}
where $y_i \in \mathcal{Y}$ denotes label, $\oplus$ denotes concatenation operator.

The overall multi level alignment and supervision object of the proposed vision-language sub-framework contains the following terms: the global sample-level alignment loss $L_\text{global}$, the local token-level alignment loss $L_\text{local}$ and the disease-level supervision loss $L_\text{disease}$. We optimize these terms jointly via a weighted sum:
\begin{equation}
    L_1 = \lambda_1 L_\text{global} + \lambda_2 L_\text{local} + \lambda_3 L_\text{disease}
\end{equation}
where $\lambda_1$, $\lambda_2$ and $\lambda_3$ are hyperparameters to balance the weights associated with each respective loss.

\vspace{-3mm}
\subsection{Multi-endoscopic Fusion}

\begin{figure}[!t]
	\centering
	\includegraphics[width=\linewidth]{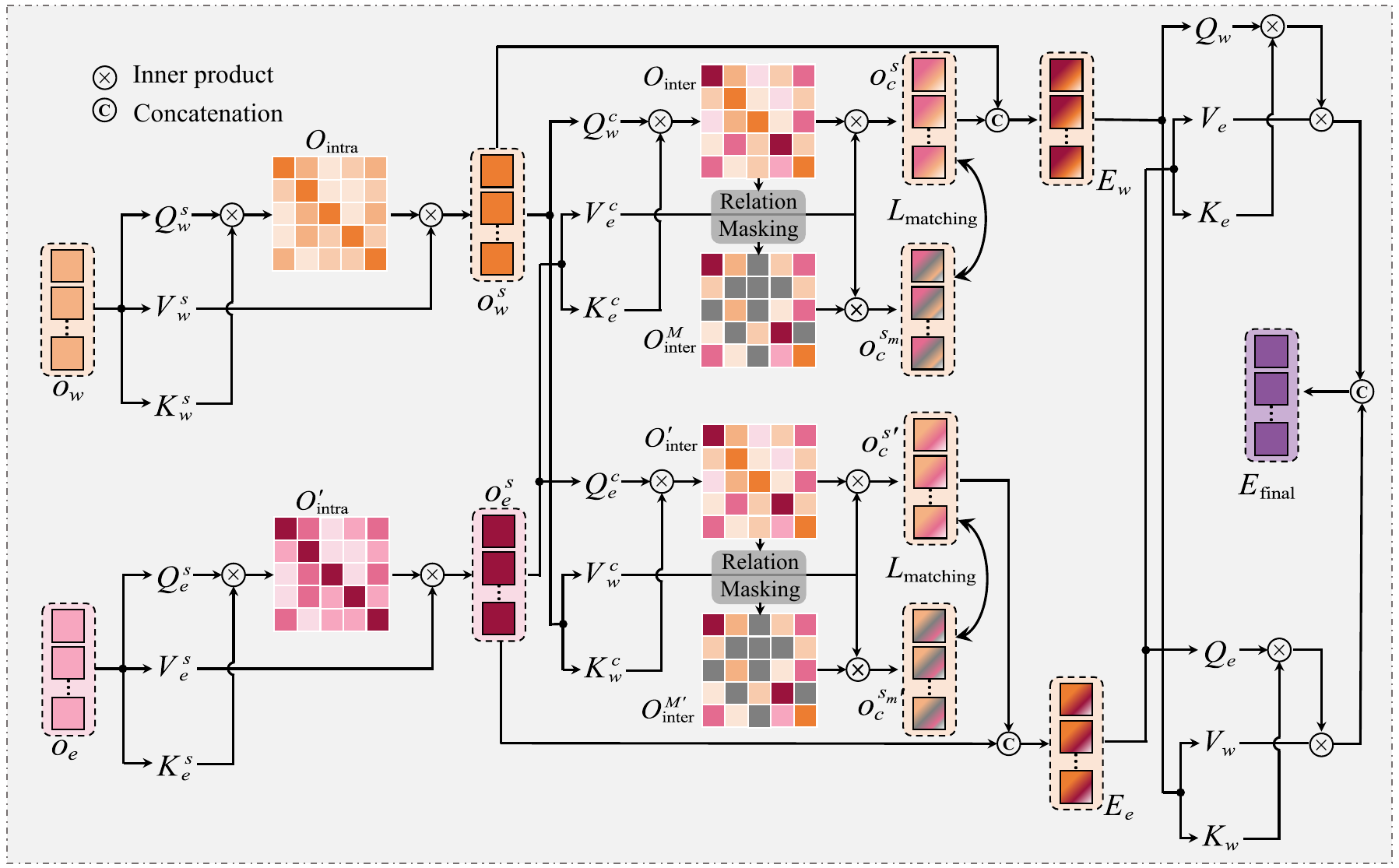}
	\caption{Detailed structure of MMRF strategy, which contains intra-modality relation learning, inter-modality relation learning and masking, and relation matching loss.}
	\label{mmrf}
        \vspace{-6mm}
\end{figure}

In the diagnosis of ulcerative colitis (UC), WLE images typically reflect surface lesions within the affected bowel segments. Characteristic WLE features include continuous, diffuse, and ulcerative lesions in the rectum and colon. Meanwhile, advanced EC techniques provide detailed histological information on mucosal, vascular, and single-cell structures. The fusion of WLE and EC facilitates a more comprehensive differential diagnosis and overall assessment of UC. Inspired by \cite{yang2023mrm}, this work proposes a multi-modal masked relation fusion (MMRF) strategy for multi-endoscopic fusion. By masking token feature relations from cross-modality perspective, MMRF enhances the representation of disease-related information. 

As shown in Fig. \ref{mmrf}, the exposition begins with a case study of the WLE branch. Given the aligned WLE image embedding $o_W$ extracted by VIT encoder in Eq.17, we apply intra-modality multi-head self attention (MSA)\cite{ilse2018attention} to generate the intra-modality relation $O_\text{intra}$ and representation $o_w^s$:
\begin{equation}
    O_\text{intra} = \text{Softmax}(\frac{Q_w^s(K_w^s)^T}{\sqrt{d}}), o_w^s = O_\text{intra}V_w^s 
\end{equation}
where $Q_w^s,K_w^s$ and $V_w^s$ are the query, key and value projected from $o_w$. $O_\text{intra}$ reflects the token-wise semantic correspondence and $o_w^s$ preserves the relative important feature representation within the WLE image embedding. Similarly, given the aligned EC image embedding $o_e$, the intra-modality relation $O_\text{intra}^{\prime}$ and representation $o_e^s$ can be calculated with the same manner.

To leverage multi-modal knowledge, a inter-modality MSA is conducted subsequent to the intra-modality MSA. Precisely, we compute inter-modality relation and representation to capture semantic correspondence between WLE and EC features:
\begin{equation}
    O_\text{inter} = \text{Softmax}(\frac{Q_w^c(K_e^c)^T}{\sqrt{d}}), o_\text{c}^s = O_\text{inter}V_e^c 
\end{equation}
where $K_e^c$ and $V_e^c$ are the key and value projected from $o_e^s$ and $Q_w^c$ is the query projected from $o_w^s$. Subsequently, in order to encourage the model to acquire both complementary multi-modal knowledge and an understanding of the relationships between modalities, we employ random masking on the original relation matrix $O_{\text{inter}}$ with a given probability $p_{\text{mask}}$ to produce $O_{\text{inter}}^M$. Then we can obtain the masked relation representation $o_c^{s_m} = O_{\text{inter}}^M V_e^c$. Similarly, given the $o_e^s$ projection as query, $o_w^s$ projections as key and value, the inter-modality relation $O_\text{inter}^{\prime}$ and representation ${o_c^s}^{\prime}$, as well as the masked relation ${o_c^s}^{\prime}$ and representation ${o_c^{s_m}}^{\prime}$ can be calculated with the same manner. To ensure the integrity of feature preservation, a relation matching loss is constructed between masked representation and complete representation as follows:
\begin{equation}
    L_{\text{matching}} = \frac{1}{N}\sum|o_c^s-o_c^{s_m}|^2+|{o_c^s}^{\prime}-{o_c^{s_m}}^{\prime}|^2
\end{equation}
Then the combined representation of WLE $E_w = o_w^s \oplus o_c^s$ and that of EC $E_e = o_e^s \oplus {o_c^s}^{\prime}$ are defined by the concatenation of each intra- and inter-modality representation. Consequently, two independent MSA are applied to generate the final fusion representation:
\begin{equation}
    E_{\text{final}} = \text{Softmax}(\frac{Q_w{(K_e)}^T}{\sqrt{d}})V_e \oplus \text{Softmax}(\frac{Q_e{(K_w)}^T}{\sqrt{d}})V_w
\end{equation}
where $Q_w, K_w, V_w$ are projected from $E_w$ and $Q_e, K_e, V_e$ are projected from $E_e$. And we define the overall loss function as:
\begin{align}
    L_\text{disease}^\prime &= - \frac{1}{N}\sum_{i=1}^Ny_i\log(F_y^{\prime}(E_\text{final}))\\
    L_2 &= \lambda_4L_\text{matching} + \lambda_5 L_\text{disease}^\prime
\end{align}
where $\lambda_4$ and $\lambda_5$ are hyperparameters. $F_y^{\prime}$ denotes a parameterized fully-connected layer.

\section{Experiments and Analysis}
\subsection{Experimental Basics}
\subsubsection{Dataset}
In this study, we retrospectively created a private dataset including WLE images, EC images, and corresponding diagnostic reports of patients with UC from the Sixth Affiliated Hospital of Sun Yat-sen University in Guangzhou, China, from March 2022 to March 2023. Notably, this study was approved by the Medical Ethics Committee of the Sixth Affiliated Hospital of Sun Yat-sen University (No. 2024ZSLYEC-391). The endoscopic examination and biopsy acquisition protocol was as follows: Using a standard colonoscope (Olympus EC 520-fold), WLE images of each colonic segment were first captured. Subsequently, staining was performed with 0.05\% crystal violet and 1\% methylene blue. Following staining, the mucosa was observed under 520-fold magnification using the colonoscope to acquire EC images. Biopsies were taken from the center of each corresponding area for histological analysis and two experienced clinicians independently scored the degree of HH using the Geboes score (GS)\cite{geboes2000GS}. The score was divided into six principal grades and four sub-grades within each grade: 0 (architectural distortion), 1 (chronic inflammatory infiltrate), 2A (eosinophilic infiltrate in the lamina propria), 2B (neutrophilic infiltrate in the lamina propria), 3 (epithelial neutrophilic infiltrate), 4 (crypt destruction), and 5 (erosions and ulcerations). A GS score below 3.2 was defined as UC HH (coded as 0), and a score of 3.2 or higher was classified as UC activity (coded as 1), serving as the outcome label for this study.

\begin{table*}[ht]
\centering
\caption{Results of our proposed VIGIL (in gray line) compared with SOTA MIL methods in UC HH prediction. For ease of identification, the top result is in bold.}
\setlength{\tabcolsep}{2mm}{
\begin{tabular}{ccccccccc}
\toprule[1.0pt]
\multirow{2}{*}{Method} & \multicolumn{4}{c}{WLE image}    & \multicolumn{4}{c}{EC image} \\  \cmidrule(r){2-5} \cmidrule(r){6-9}
             & Acc[\%]  & Sen[\%]  & Spec[\%] & AUC[\%]  & Acc[\%]  & Sen[\%] & Spec[\%] & AUC[\%]\\ \midrule
Mean-pooling & 68.84 $\pm$ 2.82  & 75.38 $\pm$ 5.75  & 66.66 $\pm$ 5.61  &74.20 $\pm$ 0.23  & 81.53 $\pm$ 4.48  & 87.69 $\pm$ 3.76   & 79.48 $\pm$ 7.06  & 84.06 $\pm$ 0.40\\
Max-pooling & 70.76 $\pm$ 4.10  & 73.84 $\pm$ 15.07  & 69.74 $\pm$ 9.37  &71.47 $\pm$ 2.21  & 79.61 $\pm$ 4.95  & 83.07 $\pm$ 8.97  & 78.46 $\pm$ 8.82  & 84.10 $\pm$ 2.48\\
Top-K pooling & 73.08 $\pm$ 2.05  & 76.92 $\pm$ 3.61  & 71.79 $\pm$ 2.92  &74.56 $\pm$ 2.43  & 82.69 $\pm$ 3.44  & 84.82 $\pm$ 4.02  & 82.05 $\pm$ 5.12  & 85.34 $\pm$ 1.59\\
ABMIL\cite{ilse2018attention} & 74.61 $\pm$ 3.72  & 64.61 $\pm$ 10.43  & 77.94 $\pm$ 7.70  &71.95 $\pm$ 0.83  & 81.15 $\pm$ 2.82  & 87.69 $\pm$ 6.15   & 78.97 $\pm$ 5.23  & 86.31 $\pm$ 1.48\\
TransMIL\cite{shao2021transmil} & 74.61 $\pm$ 3.07  & 80.00 $\pm$ 3.77  & 72.82 $\pm$ 5.27  &74.87 $\pm$ 0.59  & 83.84 $\pm$ 4.31  & 83.07 $\pm$ 8.97   & 84.10 $\pm$ 8.17  & 85.83 $\pm$ 1.65\\
FRMIL\cite{chikontwe2022frmil} & 73.07 $\pm$ 3.43  & 80.00 $\pm$ 6.15  & 70.76 $\pm$ 5.02 &74.83 $\pm$ 2.16  & 77.31 $\pm$ 2.24  & 86.15 $\pm$ 5.75  & 74.35 $\pm$ 4.58  & 83.03 $\pm$ 0.67\\
DSMIL\cite{li2021dsmil} & 70.38 $\pm$ 3.76  & 72.30 $\pm$ 3.76  & 69.74 $\pm$ 4.97  &74.20 $\pm$ 0.97  & 83.84 $\pm$ 2.30  & \textbf{89.23 $\pm$ 6.15}   & 82.05 $\pm$ 4.28  & 86.94 $\pm$ 1.75\\
PMIL\cite{yu2023prototypical} & 73.87 $\pm$ 2.61  & 78.46 $\pm$ 3.08  & 72.30 $\pm$ 2.98  &75.93 $\pm$ 2.16  & 84.23 $\pm$ 3.92  & 87.69 $\pm$ 3.76   & 83.07 $\pm$ 6.19  & 87.10 $\pm$ 0.97\\
\rowcolor{mygray}
VIGIL (KS-MIL) & 76.15 $\pm$ 4.48  & 81.54 $\pm$ 7.84  & 74.36 $\pm$ 7.06  &76.37 $\pm$ 2.14  & 86.15 $\pm$ 2.54  & 84.61 $\pm$ 4.86   & 86.66 $\pm$ 4.41  & 87.69 $\pm$ 1.19\\ \midrule
MI-Zero\cite{lu2023MIzero} & 80.02 $\pm$ 4.31  & 84.61 $\pm$ 6.88  & 78.46 $\pm$ 5.28  &80.31 $\pm$ 2.13  & 85.38 $\pm$ 1.96  & 86.15 $\pm$ 5.75   & 85.13 $\pm$ 4.41  & 87.73 $\pm$ 0.52\\
ViLa-MIL\cite{shi2024vila} & 81.92 $\pm$ 2.30  & 83.08 $\pm$ 3.08  & 81.02 $\pm$ 3.07  &81.06 $\pm$ 0.44  & 85.77 $\pm$ 0.94  & 85.64 $\pm$ 3.47  & 85.64 $\pm$ 2.05  & 88.04 $\pm$ 0.20\\
\rowcolor{mygray}
VIGIL (VLM) & \textbf{84.23 $\pm$ 1.88}  & \textbf{84.61 $\pm$ 4.86}  & \textbf{84.10 $\pm$ 3.76}  & \textbf{83.84 $\pm$ 2.61}  & \textbf{86.54 $\pm$ 2.10}  & 87.69 $\pm$ 3.76 & \textbf{86.15 $\pm$ 3.47}  & \textbf{90.84 $\pm$ 0.78}\\
\bottomrule[1.0pt]
\label{tabel1}
\vspace{-3mm}
\end{tabular}}
\end{table*}

\begin{table}[ht]
\setlength{\abovecaptionskip}{0.05cm} 
\centering
\caption{Results of our proposed VIGIL compared with SOTA multi-modal fusion methods based on endoscopic images for UC HH prediction.}
\setlength{\tabcolsep}{1mm}
\begin{tabular}{ccccc}
\toprule[1.0pt]
Method & Acc[\%]  & Sen[\%]  & Spec[\%] & AUC[\%] \\ \midrule
 SCNet\cite{ding2020scnet}   & 88.46 $\pm$ 2.10  & 83.08 $\pm$ 3.08  & 90.76 $\pm$ 1.25 & 91.61 $\pm$ 0.40\\
 Endo-CRC\cite{lu2022real}   & 90.38 $\pm$ 2.71  & 83.48 $\pm$ 2.28  & 92.82 $\pm$ 4.41 & 91.04 $\pm$ 1.52\\
 CMMCAN\cite{chong2024cmmcan} & 90.75 $\pm$ 0.78  & 84.61 $\pm$ 4.86  & 91.24 $\pm$ 1.02 & 92.78 $\pm$ 1.01\\
 \rowcolor{mygray}
 VIGIL & \textbf{92.69 $\pm$ 0.76}  & \textbf{87.69 $\pm$ 3.76}  & \textbf{95.38 $\pm$ 1.02} & \textbf{94.79 $\pm$ 0.09}\\
 \bottomrule[1.0pt]
 \label{table3}
 \vspace{-10mm}
\end{tabular}
\end{table}

This dataset comprises 87 patients with UC, including 1252 WLE images and 5401 EC images from 161 colonic segments, along with corresponding diagnostic reports. Within the dataset, 121 segments are categorized as HH and 40 as active UC. For the purposes of this study, the dataset is divided into a training set containing 90 segments, a validation set containing 19 segments and a test set containing 52 segments.

\begin{figure*}[!t]
	\centering
	\includegraphics[width=0.9\linewidth]{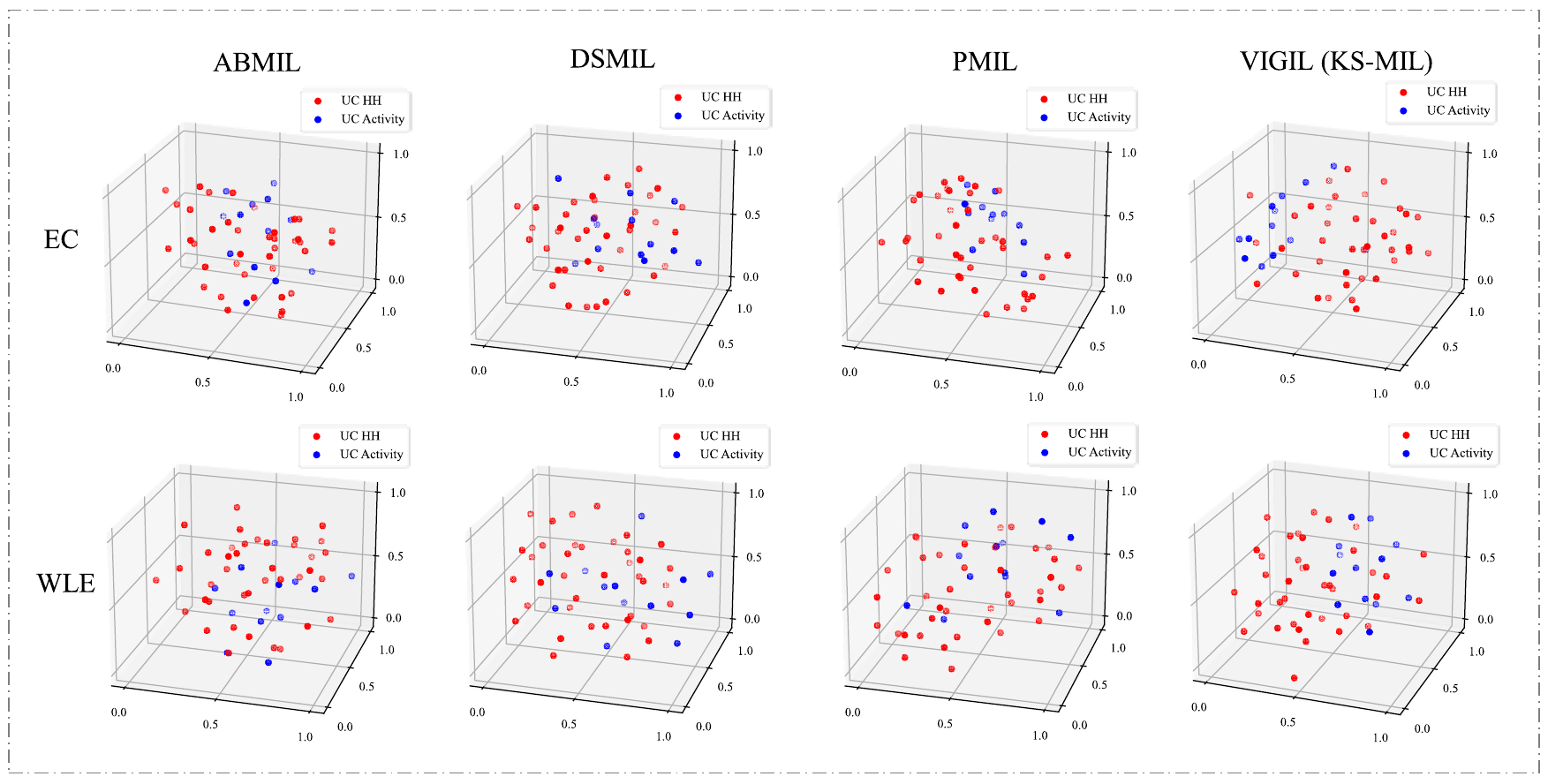}
	\caption{The t-SNE visualization of segment-level feature clustering on the EC and WLE test set. }
	\label{tsne}
        \vspace{-5mm}
\end{figure*}

\subsubsection{Evaluation Metrics}

To comprehensively evaluate the performance of different methods on the dataset, we employ the following standard classification evaluation metrics: accuracy (ACC), sensitivity (SEN), specificity (SPEC) and the area under the curve (AUC) score. The experiments are conducted five times using different random seeds, and the results presented include the average values along with their corresponding standard deviations.

\subsubsection{Implementation Details}
All experiments are conducted under the Pytorch architecture and trained on a single NVIDIA Geforce RTX 4090. In our method, the original video frames are resized to $224 \times224$ and subjected to z-score normalization as the pre-processing step. Data augmentation techniques containing rotation, flipping, and cropping are applied to alleviate overfitting during the training phase. For VIGIL architecture, a pre-trained ResNet-50\cite{he2016deep} is taken as the image encoder, while DistilBERT\cite{sanh2019distilbert} is utilized as the text encoder. Following the practice in contrastive learning\cite{radford2021clip}\cite{he2020momentum}, the projection feature dimension $d$ is 1024 and the temperature hyperparameters are $\tau_1=0.1,\tau_2=0.07$. The balanced hyperparameters $\lambda_1,\lambda_2,\lambda_3,\lambda_4,\lambda_5$ are all initialized to 1. The number of typical frames $K=3$ is the optimal value for this dataset. The optimal relation masking probability $p_{\text{mask}}$ is set to 0.15 (see Section \textit{Ablation Study} for more details). Additionally, a batch size of 32 is utilized, with the AdamW optimizer configured with a learning rate of $2\times10^{-4}$ and a weight decay of $5\times10^{-3}$. We implement a linear warmup followed by a cosine annealing scheduler, and the training process is carried out over 100 epochs.

\begin{figure*}[t]
	\centering
	\includegraphics[width=\linewidth]{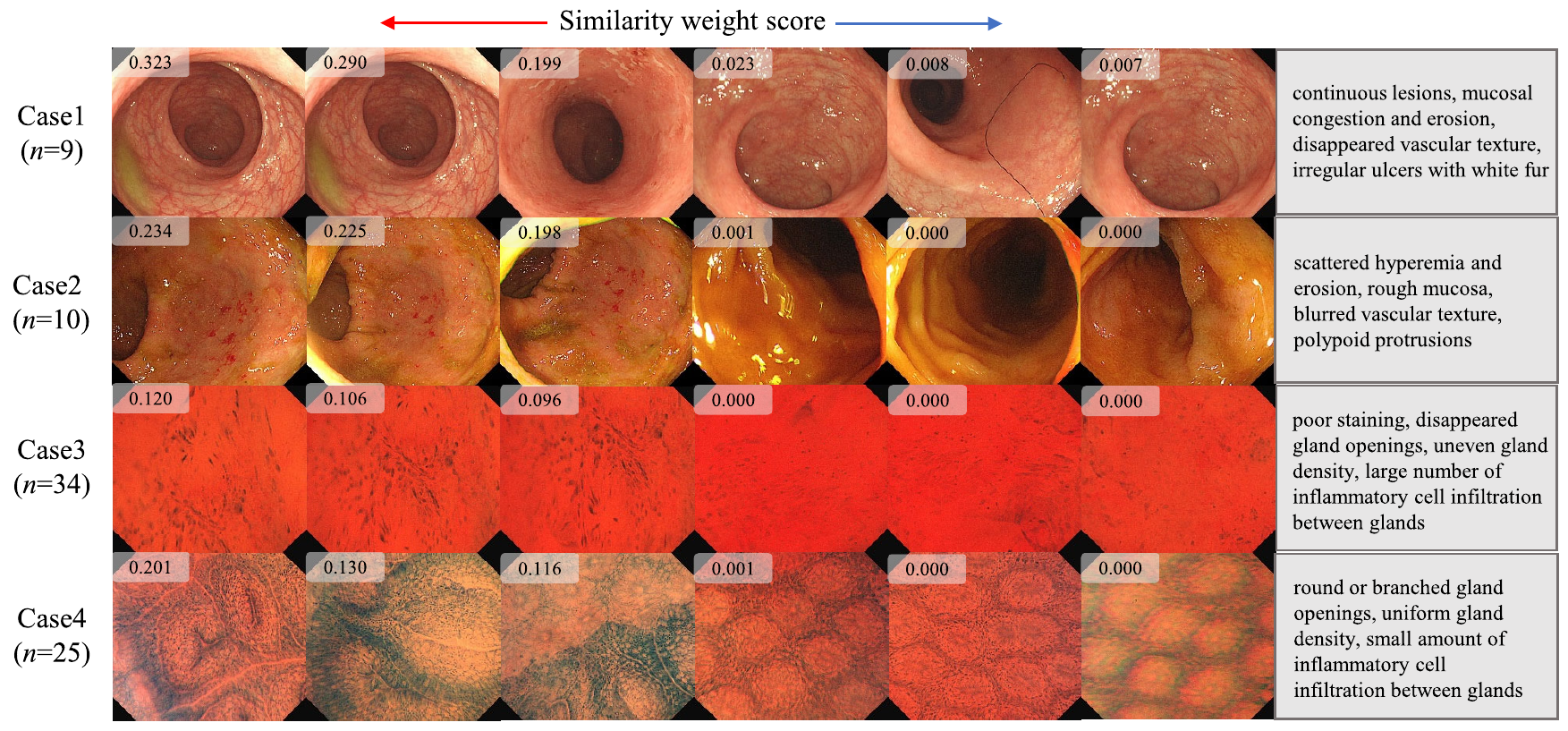}
	\caption{
    Visualization of the similarity weight scores between frames within each segment. This figure presents four cases (the number of frames $n$ is listed) from WLE and EC test set. The top three scoring frames are displayed in the first three columns, while the bottom three are shown in the last three columns. Corresponding colonic segment-level text prompts are also included.}
	\label{sim}
        \vspace{-4mm}
\end{figure*}

\vspace{-3mm}
\subsection{Performance of UC HH Prediction}
\subsubsection{Quantitative Results}

We evaluate the performance of VIGIL for colonic segment-level classification in predicting UC HH. As shown in Table \ref{tabel1}, we compare the performance of VIGIL against other SOTA MIL methods for this UC HH prediction task. To ensure a fair comparison, we evaluate the VIGIL model trained solely on image data (denoted as VIGIL (KS-MIL)) against conventional and deep-learning-based MIL methods. Furthermore, we benchmark the VIGIL model trained jointly on image and text data (denoted as VIGIL (VLM)) against SOTA VLM-based MIL approaches. Table \ref{table3} presents a performance comparison between VIGIL and various other multi-modal fusion methods based on endoscopic images. All results indicate that VIGIL demonstrates a significant advantage over existing methods in addressing the UC HH prediction task.

When compared with traditional MIL methods including Max-, Mean-, and Top-K pooling, the advantages of VIGIL become even more evident. Traditional MIL methods rely on handcrafted non-parametric operators, which limit their flexibility. In addition, relying solely on instance-level output predictions may fail to uncover deeper relationships between the bag and its instances. Specifically, Max-pooling is highly sensitive to outliers, while Mean pooling treats each instance with equal weight, which can negatively impact the final prediction results. As reported in Table \ref{table3}, our proposed VIGIL (KS-MIL) achieves a great improvement of 3.07-7.31\% in ACC and 1.81-4.9\% in AUC across WLE image and EC image.

Compared with advanced deep-learning-based MIL methods (ABMIL\cite{ilse2018attention}, TransMIL\cite{shao2021transmil}, FRMIL\cite{chikontwe2022frmil}, DSMIL\cite{li2021dsmil}, and PMIL\cite{yu2023prototypical}), VIGIL (KS-MIL) demonstrates superior performance across various metrics. In terms of AUC, our proposed method outperforms the aforementioned leading MIL methods, specifically DSMIL and PMIL, by margins of 2.17\% and 0.44\% on the WLE image dataset, and 0.75\% and 0.59\% on the EC image dataset, respectively. By introducing the KS-MIL module, VIGIL effectively facilitates the selection of key typical frames and computes the priority of each frame in relation to disease prediction using similarity weights, which improves the integration of instance features.

\begin{table*}[!ht]
\vspace{-3mm}
\centering
\caption{Ablation experimental results of different similarity metrics in KS-MIL for UC HH prediction. }
\setlength{\tabcolsep}{2mm}{
\begin{tabular}{ccccccccc}
\toprule[1.0pt]
\multirow{2}{*}{Similarity Metric} & \multicolumn{4}{c}{WLE image}    & \multicolumn{4}{c}{EC image} \\  \cmidrule(r){2-5} \cmidrule(r){6-9}
             & Acc[\%]  & Sen[\%]  & Spec[\%] & AUC[\%]  & Acc[\%]  & Sen[\%] & Spec[\%] & AUC[\%]\\ \midrule
Euclidean & 72.69 $\pm$ 2.82  & 78.46 $\pm$ 5.75  & 70.76 $\pm$ 4.75  &74.71 $\pm$ 1.29  & 84.54 $\pm$ 1.23  & 84.00 $\pm$ 3.77   & 84.71 $\pm$ 2.05  & 85.79 $\pm$ 1.10\\
Manhattan 
& 73.53 $\pm$ 2.55  & 76.92 $\pm$ 4.86  & 69.74 $\pm$ 4.97  &73.88 $\pm$ 0.47
& 83.38 $\pm$ 1.88  & 84.07 $\pm$ 6.88  & 83.97 $\pm$ 4.10  & 84.37 $\pm$ 0.85\\
KL-divergence & 74.07 $\pm$ 1.72  & 80.00 $\pm$ 6.15  & 70.77 $\pm$ 3.47  &75.74 $\pm$ 0.41  & 85.00 $\pm$ 2.55  & 83.22 $\pm$ 3.86  & 85.13 $\pm$ 4.97  &86.03 $\pm$ 0.49\\
\rowcolor{mygray}
Cosine & \textbf{76.15 $\pm$ 4.48}  & \textbf{81.54 $\pm$ 7.84}  & \textbf{74.36 $\pm$ 7.06}  &\textbf{76.37 $\pm$ 2.14}  & \textbf{86.15 $\pm$ 2.54}  & \textbf{84.61 $\pm$ 4.86}   & \textbf{86.66 $\pm$ 4.41}  & \textbf{87.69 $\pm$ 1.19}\\ 
\bottomrule[1.0pt]
\label{table2}
\vspace{-7mm}
\end{tabular}}
\end{table*}

\begin{table*}[ht]
\centering
\caption{Ablation experimental results of different components in multi-level alignment and supervision for UC HH prediction.}
\setlength{\tabcolsep}{2mm}{
\begin{tabular}{@{}ccccccccccc@{}}
\toprule[1.0pt]
\multicolumn{3}{c}{Loss} & \multicolumn{4}{c}{WLE image}    & \multicolumn{4}{c}{EC image} \\  \cmidrule(r){1-3} \cmidrule(r){4-7}  \cmidrule(r){8-11} 
 $L_{\text{global}}$ & $L_{\text{local}}$ & $L_{\text{disease}}$ & Acc[\%]  & Sen[\%]  & Spec[\%] & AUC[\%]  & Acc[\%]  & Sen[\%] & Spec[\%] & AUC[\%]\\ \midrule
- & - & \checkmark & 76.15 $\pm$ 4.48  & 81.54 $\pm$ 7.84  & 74.36 $\pm$ 7.06  &76.37 $\pm$ 2.14  & 86.15 $\pm$ 2.54  & 84.61 $\pm$ 4.86   & 86.66 $\pm$ 4.41  & 87.69 $\pm$ 1.19\\
\checkmark & - & \checkmark & 80.77 $\pm$ 1.71  & 83.23 $\pm$ 3.76  & 77.95 $\pm$ 2.05  &83.19 $\pm$ 0.50  & 85.00 $\pm$ 1.44  & 87.69 $\pm$ 3.76  & 84.10 $\pm$ 2.51  & 89.54 $\pm$ 0.59\\
- & \checkmark & \checkmark & 82.69 $\pm$ 2.71  & 83.77 $\pm$ 3.07  & 79.48 $\pm$ 3.62  &82.09 $\pm$ 0.74  & 85.92 $\pm$ 0.76  & 85.77 $\pm$ 3.07   & 85.13 $\pm$ 1.02  & 88.63 $\pm$ 0.42\\
\rowcolor{mygray}
\checkmark & \checkmark & \checkmark & \textbf{84.23 $\pm$ 1.88}  & \textbf{84.61 $\pm$ 4.86}  & \textbf{84.10 $\pm$ 3.76}  & \textbf{83.84 $\pm$ 2.61}  & \textbf{86.54 $\pm$ 2.10}  & \textbf{87.69 $\pm$ 3.76}   & \textbf{86.15 $\pm$ 3.47}  & \textbf{90.84 $\pm$ 0.78}\\
\bottomrule[1.0pt]
\label{table4}
\vspace{-5mm}
\end{tabular}}
\end{table*}

\begin{figure*}[!t]
	\centering
	\includegraphics[width=\linewidth]{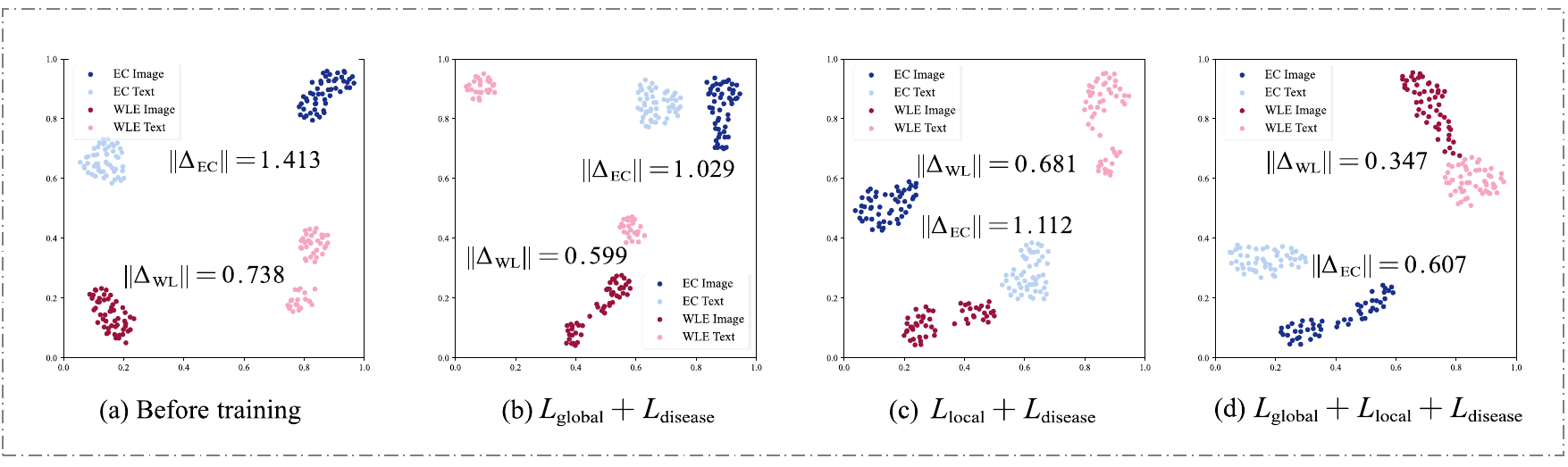}
	\caption{The UMAP visualization of image and text representations under different multi-level alignment and supervision settings. }
	\label{umap}
        \vspace{-5mm}
\end{figure*}

We also compare our VIGIL method with the advanced VLM-based methods such as MI-Zero\cite{lu2023MIzero} and ViLa-MIL\cite{shi2024vila}, while both of these methods exhibit limitations in the image-text alignment representations. They focus exclusively on contrastive learning at the global sample level when addressing the alignment, thereby lacking an exploration of local token representations. Precisely, in the WLE image dataset, MI-Zero and ViLa-MIL achieve AUC of 80.31\% and 81.06\%, respectively, while the VIGIL (VLM) attains an AUC of 83.34, representing improvements of 3.03\% and 2.28\%. In the EC image dataset, MI-Zero and ViLa-MIL both register AUC of 87.73\% and 88.04\%, respectively, whereas VIGIL (VLM) achieves an AUC of 90.84\%, marking enhancements of 3.11\% and 2.80\%. By integrating additional contextual information from the report text along with multi-level image-text alignment and supervision, our proposed method can effectively learn discriminative visual features, yielding promising classification performance.

Furthermore, a series of multi-modal fusion methods based on endoscopic images (SCNet\cite{ding2020scnet}, Endo-CRC\cite{lu2022real}, and CMMCAN\cite{chong2024cmmcan}) are selected to serve as benchmarks in our comparison experiments. However, these methods adopt simplistic operations such as concatenation, convolution, or adaptive attention mechanisms, which insufficiently capture the semantic information intrinsic to each modality as well as the complementary semantic interactions among different modalities. As illustrated in Table \ref{table3}, VIGIL demonstrates exceptional performance in the UC HH prediction task by employing a specially designed MMRF strategy, achieving an ACC of 92.69\% and an AUC of 94.79\%.

\subsubsection{Qualitative Results}
To demonstrate the effectiveness of VIGIL, we further visually compare in Fig. \ref{tsne} the features learned by our proposed method against those by the other three SOTA MIL methods, \textit{i.e.}, ABMIL\cite{ilse2018attention}, DSMIL\cite{li2021dsmil}, and PMIL\cite{yu2023prototypical}. Due to the high and varying dimensions of features across different methods, we employed t-SNE\cite{van2008tsne} to visualize segment-level feature clustering on the WLE and EC test set, reducing the dimension to three for visualization purposes. The results indicate that VIGIL (KS-MIL) yields a more compact intra-class representation and greater inter-class separation in the learned embedding space compared with other MIL methods. This improved class discrimination contributes to its superior performance and suggests the effectiveness of the proposed approach in capturing discriminative features.

For a further illustration, a series of colonic segment cases, each composed of multiple video frames, are randomly selected from WLE and EC test set. The similarity weight scores between frames within each segment are visualized in Fig. \ref{sim}. Board-certified gastroenterologists reviewed the features selected by VIGIL (those with high similarity weight scores) for each class embedding and confirmed their relevance for UC diagnosis. In WLE images, frames with high similarity weight scores exhibit features characteristic of active UC, such as mucosal bleeding, erosions, ulcers, and loss of vascular pattern, thus contributing significantly to the diagnosis. Conversely, low-scoring frames, often lacking these diagnostic features, are less informative. Similarly, in EC images, high-scoring frames correspond to histological features of UC activity, including inflammatory cell infiltration, crypt distortion, and altered crypt openings. Low-scoring frames typically lack these features and thus contribute less to the diagnostic assessment.

\vspace{-3mm}
\subsection{Ablation Study}
In this section, we conduct ablation study analysis on the effectiveness of the proposed VIGIL architecture by examining three key components containing KS-MIL, multi-level alignment and supervision, and MMRF strategy, elucidating the contributions of our approach. 

\subsubsection{Effectiveness of KS-MIL}
The core aim of KS-MIL is to improve bag-level feature representations by selecting critical typical instances and utilizing learned similarity weight metrics. In our exploration, we investigate various values of the number of typical frames $K$ and different similarity metrics to showcase the efficacy of KS-MIL. 

\begin{figure}[!t]
	\centering
	\includegraphics[width=\linewidth]{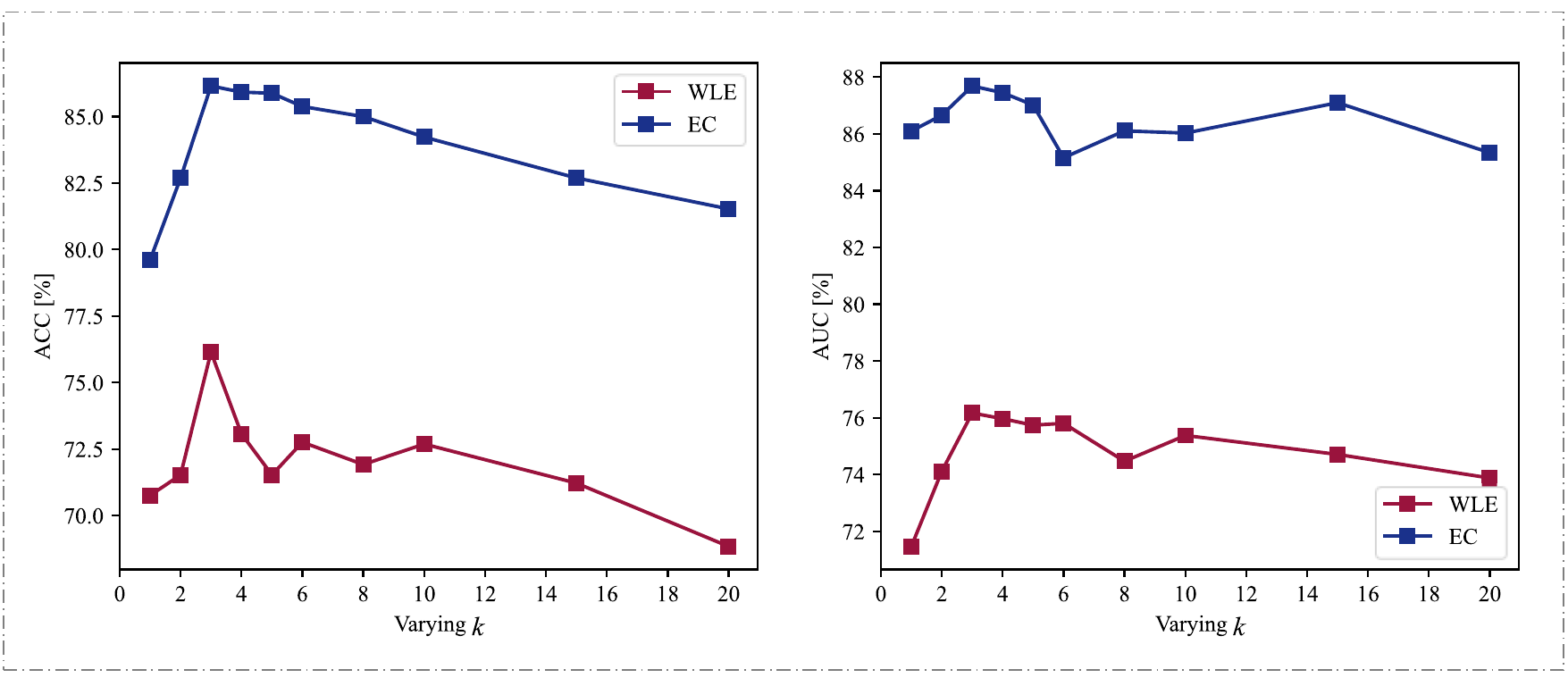}
	\caption{Results in terms of ACC and AUC obtained by varying $k$ on the WLE and EC test set.}
	\label{line}
        \vspace{-7mm}
\end{figure}

We present in Fig. \ref{line} the results for fine-tuning parameter $K$ on the WLE and EC image dataset. With the same experimental settings, we vary $K\in\{1,2,3,4,5,6,8,10,15,20\}$ to train KS-MIL and the results in terms of ACC and AUC. As observed, the experimental results achieve the best classification performance when the number of typical frames equals 3. Therefore, we finally set 3 as the default value of $K$ for our proposed method.

To determine the optimal similarity metric of our proposed KS-MIL, we conduct ablation experiments on various metrics including Euclidean, Manhattan, and KL-divergence, as shown in Table \ref{table2}. When using cosine similarity to calculate the similarity weights, it is evident that the experimental results achieve the best performance on the WLE and EC image dataset. This phenomenon can be attributed to the advantages of cosine similarity in terms of scale invariance and directional emphasis, which allows for more effective learning of the relative relationships between instances and facilitates the capture of underlying patterns and feature delineations.

\subsubsection{Effectiveness of Multi-level Alignment and Supervision}

In the field of VLM, effective alignment between images and text plays a crucial role in facilitating multimodal learning and understanding. We conduct ablation experiments on the proposed multi-level alignment and supervision with different settings. The results are shown in Table \ref{table4}, where $L_\text{global}$, $L_\text{local}$, and $L_\text{disease}$ represent the global sample-level alignment loss, the local token-level alignment loss and the disease-level supervision loss, respectively. For a fair comparison, $L_\text{disease}$ is utilized across all experiments conducted. It is observed that $L_\text{global}$ and $L_\text{local}$ contribute significantly to enhancing classification performance, indicating sample-level alignment and local-level alignment facilitate the image encoder to learn more discriminative and generalizable visual representation for classification purposes. Notably, when training $L_\text{global}$, $L_\text{local}$, and $L_\text{disease}$ jointly, the model attains the best performance.

To further analyze the impact of different settings, we visualize the segment-level feature based on UMAP\cite{sainburg2021umap} visualization with the modality gap distance, as shown in Fig. \ref{umap}. Following \cite{liang2022mind}, the modality gap $\Delta$ is defined as the difference between the center of image and text representations:
\begin{equation}
    \Delta = \frac{1}{N}\sum_{i=1}^N H_i - \frac{1}{N}\sum_{i=1}^N T_i
\end{equation}
As demonstrated in Fig. \ref{umap} (a), it is evident that the default gap between two modality embeddings is separated in both WLE and EC images before the training process, in which the modality gap is measured as $||\Delta_\text{WL}||=0.738$ and $||\Delta_\text{EC}||=1.413$, respectively. Although the incorporation of $L_{\text{global}}$ and $L_{\text{local}}$ contributes a reduction of the modality gap to some degree, it is crucial to note that the gap still exists. In Fig. \ref{umap} (d), the implementation of our proposed multi-level alignment and supervision results in a significant alleviation of the modality gap between image and text. Consequently, this gap is further reduced to $||\Delta_\text{WL}||=0.607$ and $||\Delta_\text{EC}||=0.347$.

\subsubsection{Effectiveness of MMRF}

\begin{table}[t!]
\setlength{\abovecaptionskip}{0.05cm} 
\centering
\caption{Ablation experimental results of relation masking in MMRF strategy for UC HH prediction}
\setlength{\tabcolsep}{2mm}
\begin{tabular}{ccccc}
\toprule[1.0pt]
    $p_{\text{mask}}$ & Acc[\%]  & Sen[\%]  & Spec[\%] & AUC[\%] \\ \midrule
    - & 90.69 $\pm$ 0.94  & 85.74 $\pm$ 1.91  & 91.77 $\pm$ 3.07 & 92.00 $\pm$ 0.72\\
    0.10 & 91.92 $\pm$ 1.88  & 86.66 $\pm$ 3.76  & 94.55 $\pm$ 1.25 & 93.81 $\pm$ 0.86\\
    \rowcolor{mygray}
    0.15 & \textbf{92.69 $\pm$ 0.76}  & 87.69 $\pm$ 3.76  & \textbf{95.38 $\pm$ 1.02} & \textbf{94.79 $\pm$ 0.09}\\
    0.20 & 92.08 $\pm$ 1.53  & \textbf{87.71 $\pm$ 2.05}  & 93.84 $\pm$ 3.07 & 94.01 $\pm$ 0.86\\
    0.30 & 91.38 $\pm$ 2.10  & 86.23 $\pm$ 4.08  & 93.76 $\pm$ 2.25 & 92.61 $\pm$ 1.40\\    
\bottomrule[1.0pt]
\label{table5}
\vspace{-8mm}
\end{tabular}
\end{table}

We further verify the performance of our framework with and without relation masking and explore the impact of random masking probability in the MMRF strategy. As Table \ref{table5} illustrates, integrating relation masking boosts the model classification performance, indicating the efficacy of the MMRF strategy in capturing latent disease-related information across multi-modalities. Subsequently, we investigate the impact of the random masking probability $p_\text{mask}$, a critical parameter in relation masking, on the model's performance. The results, presented in Table \ref{table5}, both overly high and overly low $p_\text{mask}$ can significantly impact the model's ability to learn disease-related information and generalize effectively. Specifically, a high masking ratio may lead to the loss of critical discriminative features and hinder the attention mechanism's ability to capture relationships between modalities, compromising feature fusion quality. In contrast, a low masking probability may cause the model to rely on memorizing visible portions, limiting its use of contextual information. Thus, a default value of 0.15 is considered appropriate $p_\text{mask}$ for the proposed method.

\section{Conclusion}
In this study, we propose a novel vision-language guided MIL framework, namely VIGIL, to achieve accurate, objective, and non-invasive UC HH prediction. In the VIGIL model, we first design a novel KS-MIL module to explore the deep latent relationships between individual video frames and typical frames, thereby enhancing the feature integration at the colonic segment level. By incorporating textual information from diagnostic reports and a specifically designed multi-level alignment and supervision between image-text pairs, our method effectively captures more discriminative and richer semantic information related to UC disease. Furthermore, we present the MMRF strategy to fuse features from two endoscopic modalities, WLE and EC, integrating representations of surface lesions and cellular-level changes for more accurate and effective UC HH prediction. Extensive experiments demonstrate the superior performance of VIGIL, outperforming existing SOTA methods in both quantitative and qualitative evaluations. Ablation studies further validate the effectiveness of individual components as well as the rationality of the chosen architecture and hyperparameters. Future work will focus on collecting more multi-center data and UC diagnostic report generation to enhance the model's applicability and explainability in real-world clinical settings, ensuring its practical utility.

\bibliographystyle{ieeetr}
\bibliography{arxiv.bbl}

\end{document}